\documentclass[12pt, a4paper]{article}
\usepackage{graphicx}
\usepackage{epsfig}
\usepackage{jcappub}
\def \lleq {\lower0.9ex\hbox{ $\buildrel < \over \sim$} ~}
\def \ggeq {\lower0.9ex\hbox{ $\buildrel > \over \sim$} ~}
\def \atridot{\stackrel{...}{a}}
\def \lam    {\Lambda}

\newcommand{\rd}{{\rm d}}

\def \om   {\Omega_m}

\def \omm  {\Omega_{0 {\rm m}}}

\def \beq  {\begin{equation}}
\def \eeq  {\end{equation}}
\def \ber  {\begin{eqnarray}}
\def \eer  {\end{eqnarray}}
\def\statei{\lbrace r,s \rbrace}
\def\statej {\lbrace r,q \rbrace}
\def \jetpl {JETP Lett.\ }
\def\mn{{Mon.\@ Not.\@ Roy.\@ Ast.\@ Soc.\ }}
\def\plb {{Phys.\@ Lett.\@ B\ }}
\newcommand{\be}{\begin{equation}}
\newcommand{\ee}{\end{equation}}
\newcommand{\ba}{\begin{eqnarray}}
\newcommand{\ea}{\end{eqnarray}}
\newcommand{\bea}{\begin{eqnarray*}}
\newcommand{\eea}{\end{eqnarray*}}

\title{Statefinder hierarchy of bimetric and galileon
models for concordance cosmology}
\author{R. Myrzakulov $^1$, M. Shahalam $^2$}
\affiliation{$^1$Department of General and Theoretical Physics, Eurasian
National University, Astana, Kazakhstan}
\affiliation{$^2$Center For Theoretical Physics, Jamia Millia Islamia, New Delhi-110025, India}

\emailAdd{rmyrzakulov@gmail.com,  mdshahalam@ctp-jamia.res.in}

\abstract{
In this paper, we use Statefinder hierarchy method to distinguish between
bimetric theory of massive gravity, galileon modified gravity and DGP models applied to late time expansion of the universe. We also carry out comparison between bimetric  and DGP models using Statefinder pairs $\statei$ and $\statej$. We show that statefinder diagnostic can differentiate between $\Lambda$CDM and above mentioned cosmological models of dark energy, and finally show that Statefinder $S_2$ is an excellent discriminant of $\Lambda$CDM and modified gravity models.  }
\date{\today}


\begin{document}

\maketitle

\section{Introduction}
There exists a belief that the observed late time cosmic acceleration is
driven by some unknown exotic energy component characterized by
negative pressure dubbed  `{\it dark energy'}
\cite{perlmutter,riess,Komatsu,review}. This hypothesis is
supported by a number of observational results related to Type Ia supernovae \cite{perlmutter,riess}, cosmic microwave background radiation and large scale structures \cite{Komatsu}.

The simplest candidate of dark energy is cosmological constant $\Lambda$ with  $p=-\rho$. There is also a variety of dynamical dark energy models
which can fit into the observations. In view of the forthcoming observations, it is of utmost importance to find ways to distinguish these models. Different diagnostic measures have been proposed in the literature to distinguish
dark energy models; $Om$, $Om$3, \cite{Om,Om3} and statefinder  diagnostics
\cite{statefinder, statefinder2} are the examples of such diagnostic measures (see also Ref. {\cite{state} on the related theme). In this paper we shall employ statefinder diagnostic to distinguish some recently proposed models of dark energy. The statefinder pair $\statei$ is a geometrical diagnostic which is algebraically related to the higher derivative of scale factor ``{\it a''} with respect to time. The deceleration parameter (q), statefinder (r) and snap ($\alpha_4$) \cite{visser} contains second, third and fourth order derivative of scale factor respectively. It is really interesting that statefinders can successfully differentiate between a large variety of dark energy models \cite{statefinder2, MSA}.

Recently, a more refined  diagnostic known as `{\it Statefinder hierarchy, $S_{n}$}' is introduced in Ref. \cite{maryam}.
The statefinder diagnostics for nonminimally coupled scalar field system
and galileon field which is generically nonminimally coupled system, has
been investigated in Ref. \cite{MSA}; in which we have shown that the
nonminimally coupled scalar field and galileon models are successfully
differentiated from other popular dark energy models such as chaplygin
gas, quintessence and Dvali, Gabadadze and Porrati (DGP) models in {\it r - s} and {\it r - q} plane. DGP is plagued with ghost problem and it is also not supported by data \cite{alamg1} (see also Ref. {\cite{brane} on the related theme).


In the present paper we consider bimetric (bigravity) model of massive
gravity  which is closely related to
nonlinear massive gravity {\it a la} $dRGT$ \cite{dRGT,sdmg,Hassan,akrami}. An interesting scalar field dubbed galileon, despite the higher order derivative terms does not suffer from Ostrogradki's ghosts. Galileons emerge in $dRGT$ in the so called decoupling limit which is a valid limit for studying the Vainshtein mechanism and galileon is a natural device to
implement the latter. In fact, the lower  order galileon lagrangian is responsible for the consistency of DGP with local physics. A large number of papers are devoted to cosmological dynamics of galileon field. It was first demonstrated in Ref. \cite{Gannouji:2010au} that one needs a higher order galileon system to achieve de Sitter solution.

It is interesting to distinguish the bimetric model from the models based on galileon field. In this paper, we use statefinder pairs $\statei$ and $\statej$ to differentiate between bimetric and DGP models; we also use statefinder hierarchy $S_{n}$ to dicriminate between $\Lambda$CDM and modified gravity models of dark energy.

\section{The Statefinder Hierarchy }
\vspace{3mm}

In, what follows; we shall work in the framework of
spatially homogeneous and isotropic universe; in this case, the
scale factor $a(t)$ is the only dynamical variable.
 Since we shall be interested in the late time behavior of expansion of the universe, we consider the  Taylor expansion of the scale factor around the
present epoch $t_0$ \cite{maryam}:
\beq a(t) = a(t_{0}) + a(t_{0})\sum^{\infty}_{n=1}
\frac{\alpha_{n}(t_{0})}{n!}\left[H_{0}(t-t_{0})\right]^n,
\label{scale} \eeq
where, $ \alpha_n=\frac{d^{n}a}{dt^{n}} / {(aH^n)}$ ;
$a^{(n)}$ is the $n^{th}$ derivative of the scale factor with respect to time. The deceleration parameter is defined as
\beq\label{eq:a2} q = -\frac{\ddot a}{aH^2} = - \alpha_2 \equiv -\frac{\dot H}{H^2} - 1
\eeq
The statefinder pair $\statei$ and the Snap are defined as

\ber\label{eq:a3}
r= \frac{\atridot}{a H^3} = \alpha_3 &\equiv& \frac{\ddot H}{H^3} +  3\frac{\dot H}{H^2} + 1\,\,,\\
s= \frac{r-1}{3(q-1/2)},\,\,
\eer
\ber \label{eq:a4} \alpha_4 =
\frac{\ddddot a}{a H^4}  &\equiv& 1 + \frac{\dddot H}{H^4} +
4\frac{\ddot H}{H^3} + 3\frac{\dot H^2}{H^4} + 6\frac{\dot H}{H^2}.
\eer
The Statefinder hierarchy $S_n$ is given by Ref. \cite{maryam}:
\begin{eqnarray}
&&S_2 := \alpha_2 + \frac{3}{2}\om,\\
&&S_3  := \alpha_3,\\
&&S_4 := \alpha_4+ \frac{3^2}{2}\om ~~~~\mbox{and~so~on},
\label{state0}
\end{eqnarray}
where $\alpha_2$, $\alpha_3$, $\alpha_4$ are given by (\ref{eq:a2}), (\ref{eq:a3}), (\ref{eq:a4});
$\Omega_m = \omm (1+z)^3/h^2(z)$ and $\Omega_m = \frac{2}{3}(1+q)$ in concordance cosmology ($\Lambda$CDM). It is remarkable to see that for concordance cosmology  $S_n=1$, for $n\geq2$ during the entire history of the universe.
\\
For $n>3$, there is another way to define a null diagnostic. Using the relationship $\Omega_m = \frac{2}{3}(1+q)$ which is valid in concordance cosmology, the alternate form of the Statefinder is given as follows:
\ber\label{eq:s3s4}
&&S_4^{(1)} := \alpha_4 + 3(1+q),
\eer
for $\Lambda$CDM, $S_4^{(1)} := S_4 = 1$, and for other dark energy models  the two definitions  $ S_4$ and $S_4^{(1)} $ give different results, as demonstrated in figure \ref{s4z41znew}.
\\
The second Statefinder corresponding to $S_3^{(1)} := S_3$ is defined as follows:
\beq
S_3^{(2)} = \frac{S_3^{(1)}-1}{3(q-1/2)}~,
\label{eq:second_state_s3}
\eeq
consequently, $\lbrace S_3^{(1)},S_3^{(2)}\rbrace = \lbrace 1, 0\rbrace$ for $\Lambda$CDM. Similarly the second Statefinder corresponding to (\ref{eq:s3s4}) is defined as:
\beq
S_4^{(2)} = \frac{S_4^{(1)}-1}{9(q-1/2)}~,
\label{eq:second_state_s4}
\eeq
for $\Lambda$CDM, $\lbrace S_4^{(1)},S_4^{(2)}\rbrace = \lbrace 1, 0\rbrace$.
\begin{figure*} \centering
\begin{center}
$\begin{array}{c@{\hspace{0.4in}}c}
\multicolumn{1}{l}{\mbox{}} &
        \multicolumn{1}{l}{\mbox{}} \\ [0.cm]
\epsfxsize=2.5in
\epsffile{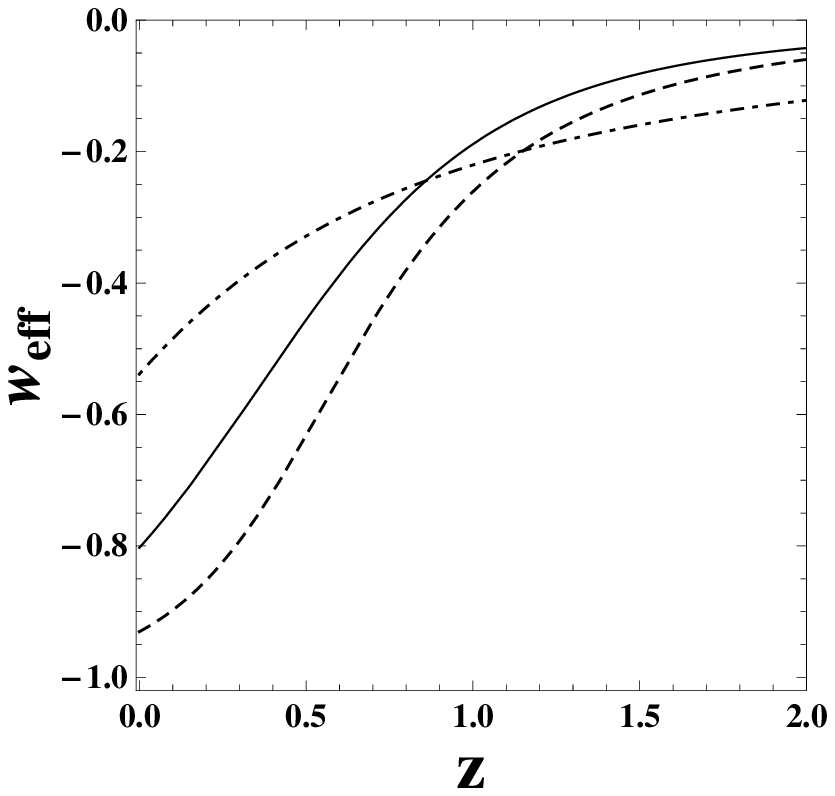} &
        \epsfxsize=2.45in
        \epsffile{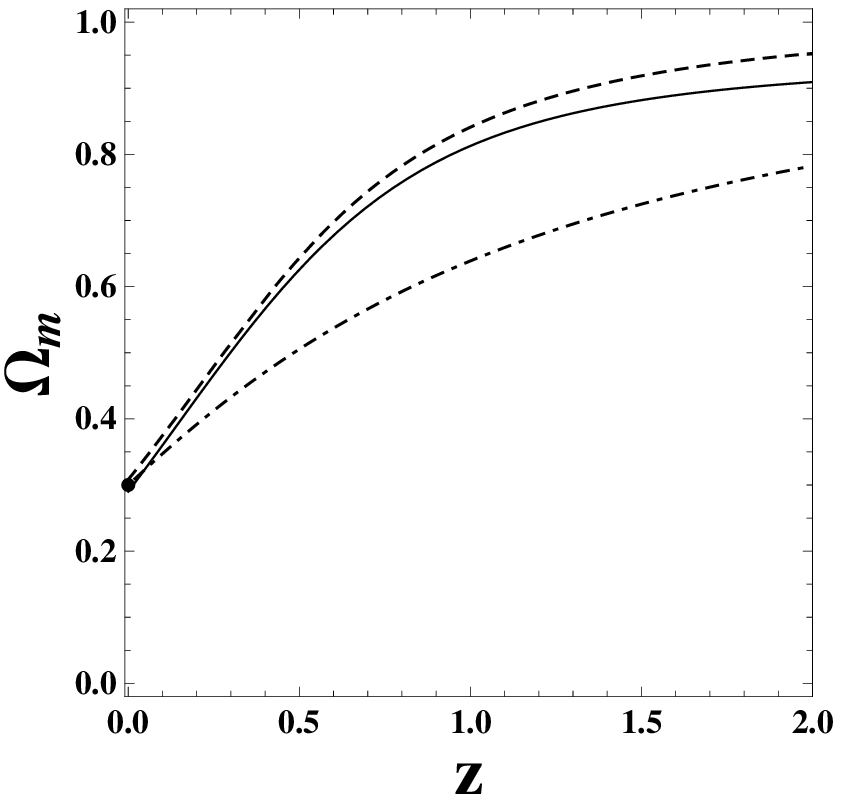} \\ [0.20cm]
\mbox{\bf (a)} & \mbox{\bf (b)}
\end{array}$
\end{center}
\caption{ \small The left (right) panel shows the evolution of $w_{eff}$ ($\Omega_{m}$) plotted against redshift z. The dark energy models are: bimetric (dashed line), galileon (solid line) and DGP (dotdashed line).
}\label{weff}
\end{figure*}
\begin{figure*} \centering
\begin{center}
$\begin{array}{c@{\hspace{0.4in}}c}
\multicolumn{1}{l}{\mbox{}} &
        \multicolumn{1}{l}{\mbox{}} \\ [0.0cm]
\epsfxsize=2.5in
\epsffile{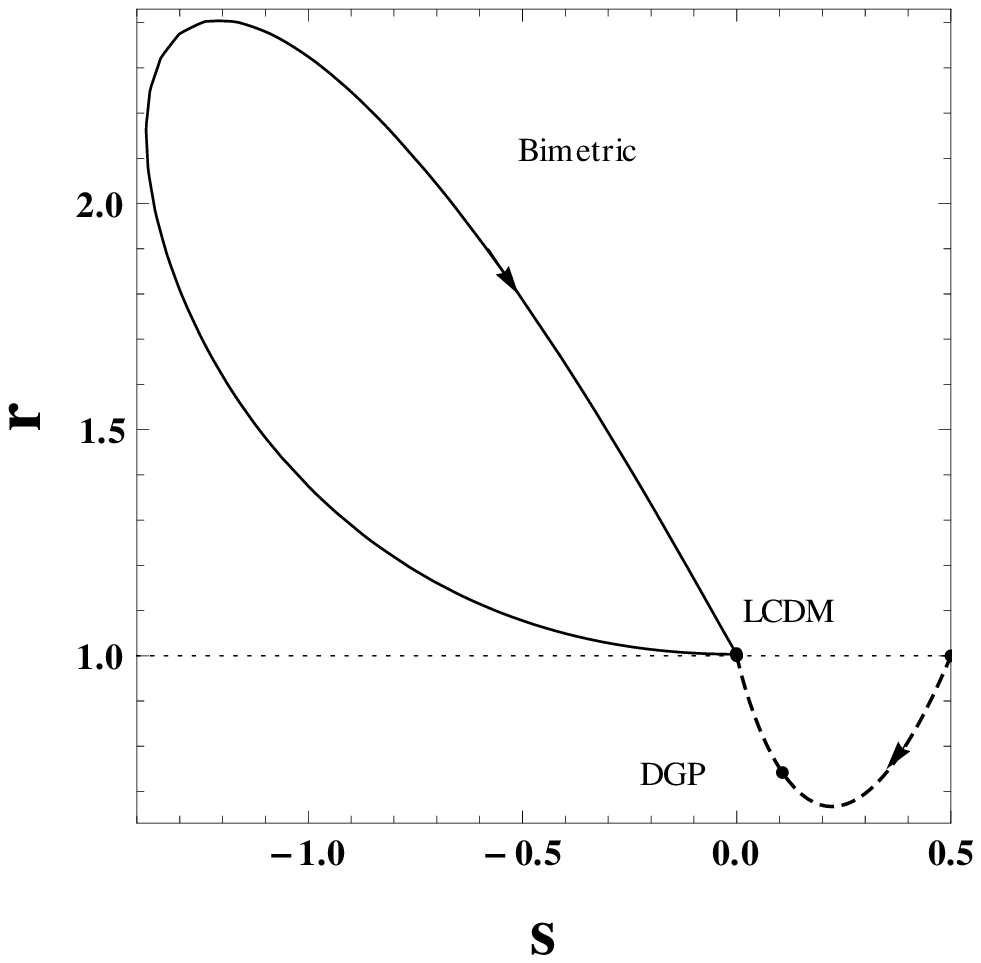} &
        \epsfxsize=2.5in
        \epsffile{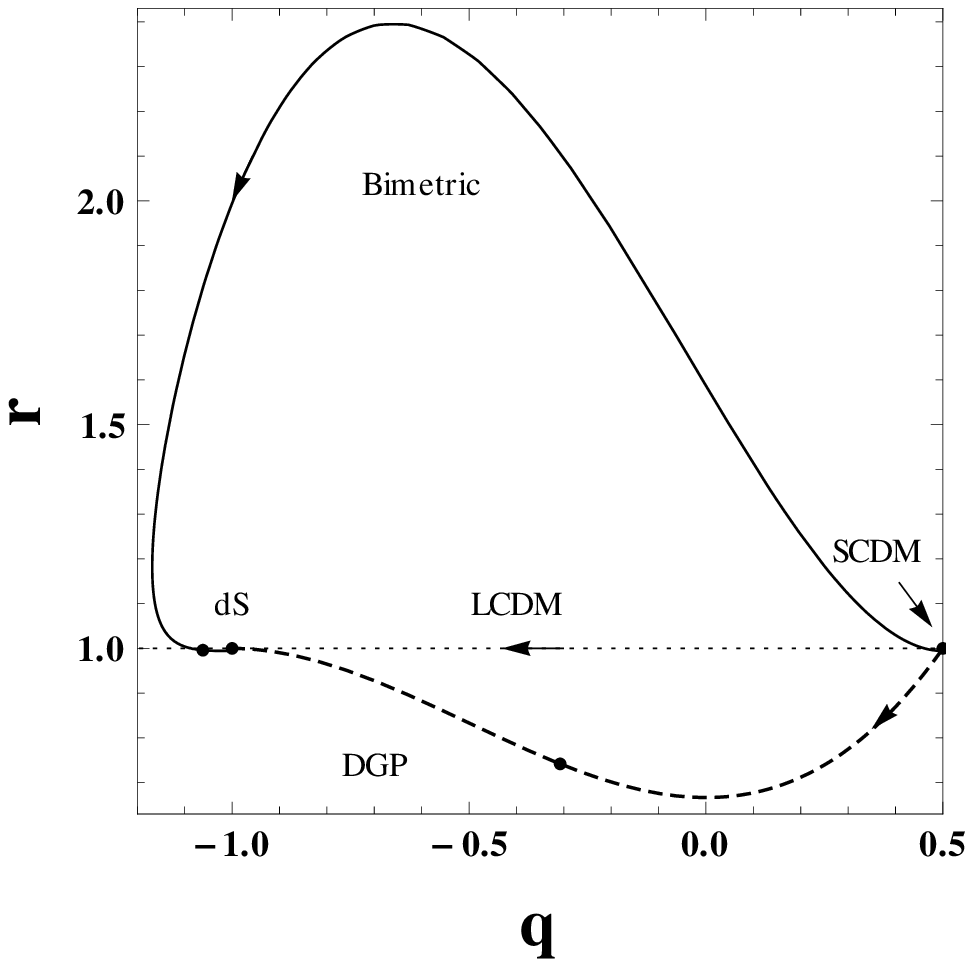} \\ [0.20cm]
\mbox{\bf (a)} & \mbox{\bf (b)}
\end{array}$
\end{center}
\caption{ \small  The panel (a)  shows the time evolution of the statefinder pair $\lbrace r,s \rbrace$ for bimetric (silod line) and DGP (dashed line) models. Bimetric model lies to the left of the LCDM fixed point ($r=1,s=0$) and DGP model lies to the right of the LCDM fixed point ($r=1,s=0$). For Bimetric model, $s$ remains at zero, whereas $r$ first increases from unity to a maximum value, then decreases to unity. For DGP model, $s$ decreases to zero, whereas $r$ first decreases from unity to a minimum value, then increases to unity. Both models converge to the fixed point ($r=1, s=0$) which corresponds to LCDM. The panel (b) shows the time evolution of the statefinder pair $\lbrace r,q\rbrace$ for bimetric (silod line) and DGP (dashed line) models. Both models diverge at the same point ($r=1,q=0.5$) which corresponds to a matter dominated universe (SCDM) and converge to the same point ($r=1,q=-1$) which corresponds to the de Sitter expansion (dS). The dark dots on the curves show current values $\lbrace r_0, s_0\rbrace$ (left) and $\lbrace r_0, q_0\rbrace$ (right).
 In both models, $\omm = 0.3$ at the current epoch.}
\label{rs-rq}
\end{figure*}
\begin{figure*} \centering
\begin{center}
$\begin{array}{c@{\hspace{0.4in}}c}
\multicolumn{1}{l}{\mbox{}} &
        \multicolumn{1}{l}{\mbox{}} \\ [0.0cm]
\epsfxsize=2.5in
\epsffile{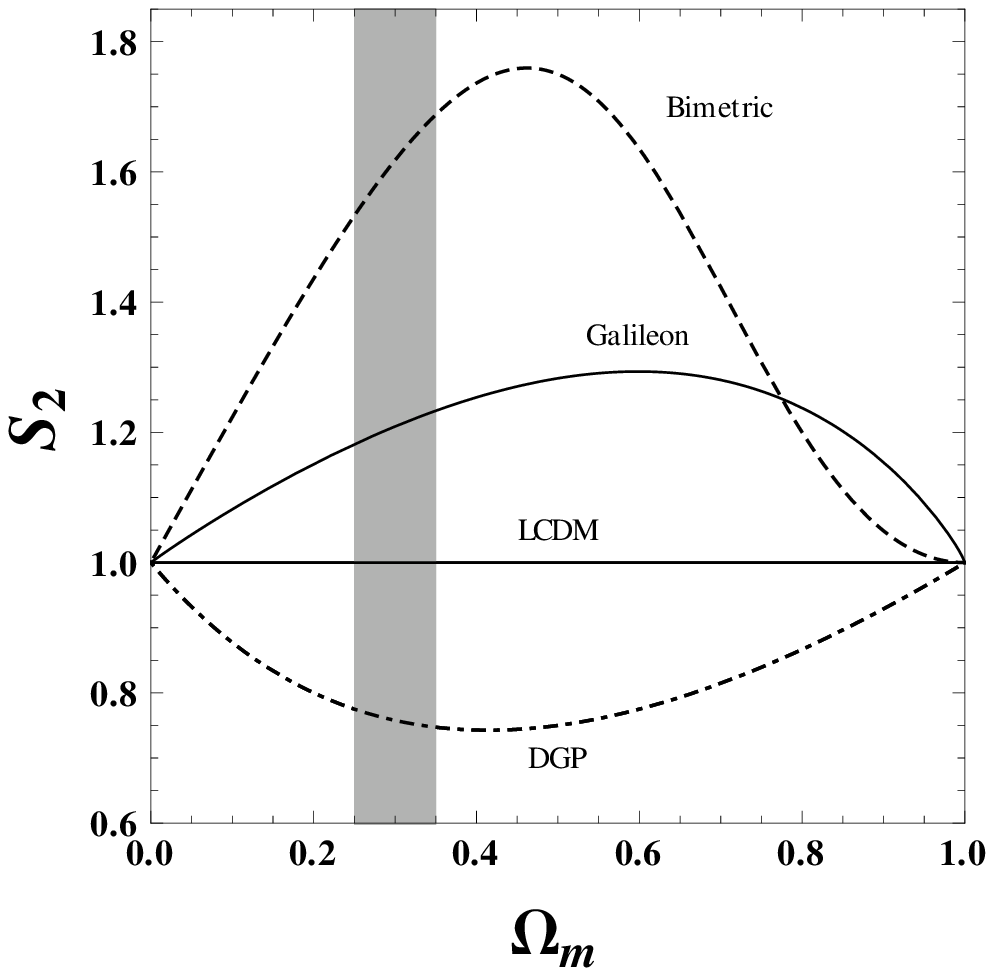} &
        \epsfxsize=2.44in
        \epsffile{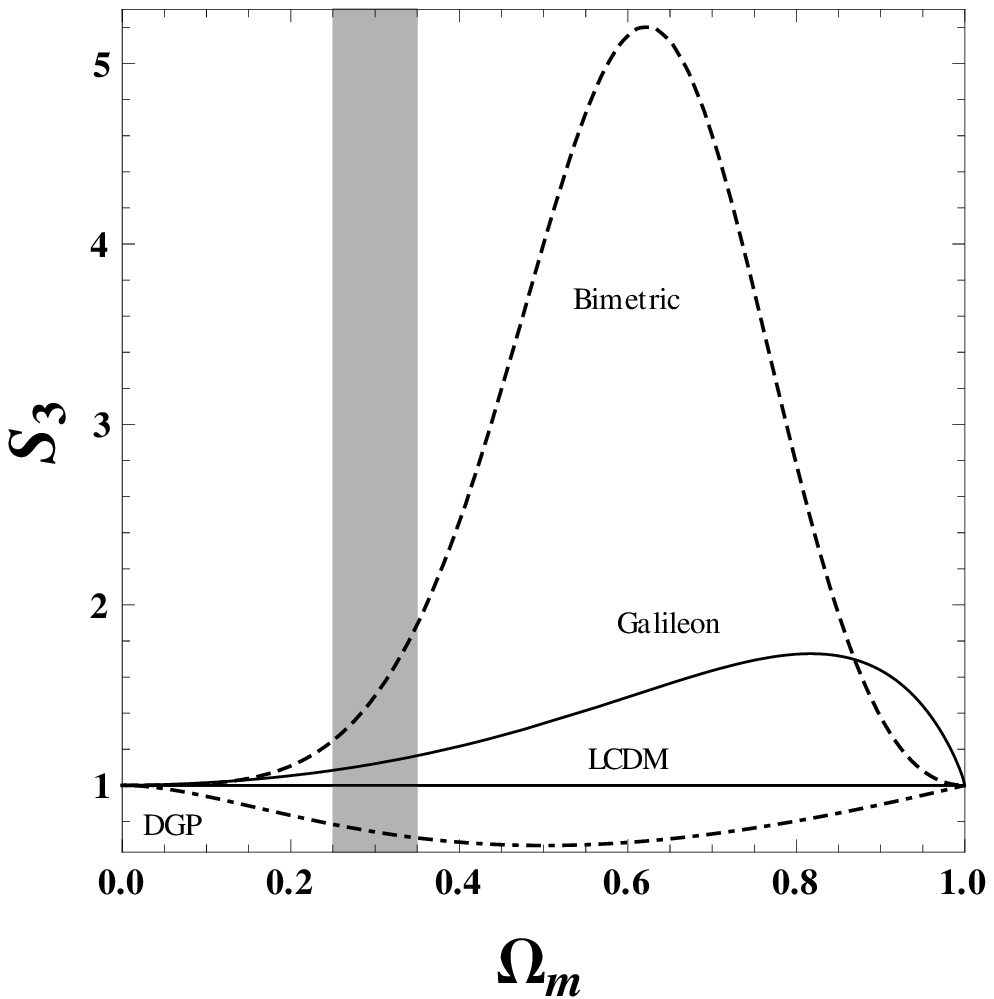} \\ [0.20cm]
\mbox{\bf (a)} & \mbox{\bf (b)}
\end{array}$
\end{center}
\caption{ \small This figure shows the evolution of
the Statefinder $S_2 \equiv {\ddot a}/{aH^2} + \frac{3}{2}\om$ and $S_3^{(1)} = S_3 \equiv {\atridot}/{a H^3}$ plotted against $\Omega_m = \omm (1+z)^3/h^2(z)$. Large values $\om \to 1$ correspond to the distant
past ($z \gg 1$), while small values $\om \to 0$ correspond to the
future ($z \to -1$).
The vertical band centered at $\omm = 0.3$ roughly corresponds to the
present epoch. In agreement with the results of Ref. \cite{Gannouji:2010au}, we used the values of model parameters in (\ref{actiong}): $\beta=0.1, c_2=1, \tilde{c}_3=15(\tilde{c}_3=c_3H_0^-2)$ and $\tilde{c}_4=4(\tilde{c}_4=c_4H_0^-4)$. With these parameters, we have equation of state parameter $ \omega \simeq -1.1$ consistent with WMAP results.} \label{s2s31}
\end{figure*}
\section{Dark Energy Models}
\vspace{3mm}

\begin{itemize}
\item {\bf Bimetric theory of massive gravity}
following Refs. \cite{ Hassan,akrami}, we consider the bimetric
massive gravity action
\begin{eqnarray}
S &=& - \frac{M^2_g}{2}\int d^4x\sqrt{-\det g}R(g) - \frac{M^2_f}{2}\int
d^4x\sqrt{-\det f}R(f) \nonumber\\
  && + m^2M_g^2\int d^4x\sqrt{-\det
g}\sum_{n=0}^{4}\beta_ne_n\left(\sqrt{g^{-1}f}\right)  \nonumber\\
&& + \int d^4x\sqrt{-\det
g}\mathcal{L}_m\left(g, \Phi\right)\label{eq:ActionOriginal} .
\end{eqnarray}
This theory contains two space-time metrics, {\it g} and {\it f}. The {\it g} metric is assumed to be a physical metric, and the {\it f}  metric is a reference metric. The theory is ghost free, and reproduces $dRGT$ in a certain limit; the generalized Friedmann equations for a flat universe ($\kappa=0$) are given by \cite{akrami}:
\begin{eqnarray}\label{eq:firstFriedmanng}
3\left(\frac{\dot{a}}{a}\right)^2-m^2\left[\beta_0+3\beta_1\frac{Y}{a}+3\beta_2\frac{Y^2}{a^2}+\beta_3\frac{Y^3}{a^3}\right]
=\frac{1}{M_g^2}\left(\rho_m+\rho_r \right),
\end{eqnarray}
\begin{eqnarray}\label{eq:spacialFieldEqg}
  -2\frac{\ddot{a}}{a}-\left(\frac{\dot{a}}{a}\right)^2 +
 m^2\left[\beta_0 + \beta_1\left(2\frac{Y}{a} + \frac{\dot{Y}}{\dot{a}}\right) \nonumber \right. \\
 \left. +\beta_2\left(\frac{Y^2}{a^2} + 2\frac{Y\dot{Y}}{a\dot{a}}\right)   + \beta_3 \frac{Y^2\dot{Y}}{a^2\dot{a}}\right] = \frac{1}{3M_g^2}\rho_r,
\end{eqnarray}
\begin{equation}\label{eq:firstFriedmannf}
3\left(\frac{\dot{a}}{Y}\right)^2- m^2\left[\beta_4 + 3\beta_3\frac{a}{Y} +
3\beta_2\frac{a^2}{Y^2} + \beta_1\frac{a^3}{Y^3}\right] = 0,
\end{equation}
\begin{eqnarray}\label{eq:spacialFieldEqf}
m^2\left[\beta_4 + \beta_3\left(2\frac{a}{Y} +
\frac{\dot{a}}{\dot{Y}}\right) + \beta_2\left(\frac{a^2}{Y^2} +
2\frac{a\dot{a}}{Y\dot{Y}}\right) \nonumber \right. \\
 \left.  + \beta_1
\frac{a^2\dot{a}}{Y^2\dot{Y}}\right]
-2\frac{\ddot{a}\dot{a}}{Y\dot{Y}}-\left(\frac{\dot{a}}{Y}\right)^2=0,
\end{eqnarray}
where, $\rho_m$  and $\rho_r$ are the energy densities of matter and
radiation respectively.
The Hubble parameter and $w_{eff}$ for this model have the form
\begin{eqnarray}
\label{eq:bm}
H(z) = {H_0} \left\lbrack \omm(1+z)^3 + \frac{B_0}{3} + B_1y  + B_2y^2  + \frac{B_3}{3}y^3 \right\rbrack^{1/2},
\end{eqnarray}
\begin{eqnarray}
\label{eq:bmweff}
w_{eff}&=&  -  \frac{\left( B_1 y' + 2 B_2 y y'+ B_3 y^2 y' - 3\Omega_{0m} a^{-3}\right)}
{ \left(B_0 + 3 B_1  y + 3 B_2  y^2 + B_3 y^3 +3\Omega_{0m} a^{-3} \right)}-1.
\end{eqnarray}
where,  $B_i \equiv m^2\beta_i/H_0^2$,  $\beta_i$  can be
absorbed into $m^2$;    $Y(Y=ya)$ , $a$ are the scale factors
corresponding to the metric $f$ and $g$ respectively and  $'$
denotes derivative with respect to $lna$. As for $m$, it should be
of the order of $H_0$ to be relevant to late time cosmic acceleration, thereby $B_i\sim 1$. While carrying out the statefinder analysis, we shall make a convenient choice, $B_i=1.44$ \cite{akrami}.

\begin{figure}
   \begin{center}
     \includegraphics[scale=0.60]{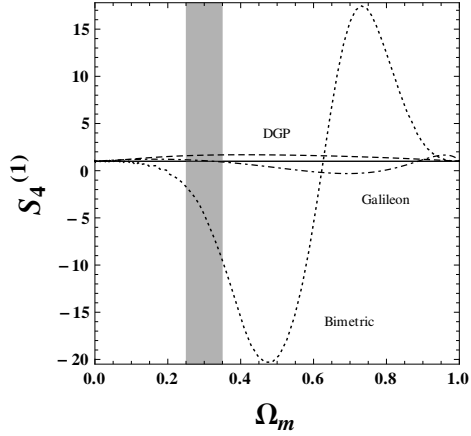}
 \end{center}
 \caption{ \small This figure shows the evolution of the Statefinder $S_4^{(1)}$ plotted
against $\om $.
} \label{s41}
\end{figure}
\item {\bf Galileon  model} \cite{sdg, Gannouji:2010au, Galileon, DeFelice, linder}:

The Galileon is a massless scalar field ($\pi$); whose action is invariant under the Galilean transformation~ $\pi(x)\rightarrow \pi(x)+ b_{\mu} x^{\mu}+ c$, where $b_{\mu}$ and $c$ are the constant four vector and scalar respectively. Following Ref. \cite{Gannouji:2010au}, we consider the galileon action
\be
\mathcal{S}=\int{\textit{\rd}^4 x}\sqrt{-g}
\left(\frac{R}{2}+c_\textit{i}~L^{(\textit{i})}\right)+\mathcal{S}_m[\psi_m,e^{2\beta\pi}g_{\mu\nu}],
\label{actiong} \ee
where, ${c_i^{'s}}$ are  constants, $\beta$ is coupling of field with
matter and  $L_i^{'s}$ \cite{Gannouji:2010au} are Lagrangians for
Galileon field.  $L_1$ is linear in field and is often omitted
assuming $c_1=0$, $L_2$ represents the standard kinetic term,
$L_3=(\partial_\mu \pi)^2 \Box \pi$ is the third order galileon term
which occurs in DGP; $L_4$ and $L_5$ are higher order Lagrangians. Following Ref. \cite{Gannouji:2010au}, we shall
consider Galileon Lagrangian upto $L_4$, sufficient to obtain a
stable de Sitter solution. As for the parameters ${c_i^{'s}}$ and
$\beta$, we adopt the choice of the said reference which gives rise
to a viable late time cosmology. It is interesting to note that the
observational constraint on $\beta$($\beta \leq 0.1$) found in
Ref. \cite{Gannouji:2010au} was independently confirmed by  the
observed limit on the time variation of Newtonian constant $G$
\cite{babichev}. Following this development, we shall use the
numerical value of $\beta=0.1$ in the statefinder analysis.

In this case, the evolution equations in a spatially flat background
have the form \cite{Gannouji:2010au}:
\begin{eqnarray}
\label{eq:Fried1}
3H^2&=&\rho_m+\frac{c_2}{2}\dot{\pi}^2-3c_3 H\dot{\pi}^3+\frac{45}{2}c_4 H^2\dot{\pi}^4,\\
\label{eq:Fried2}
2\dot{H}+3H^2&=&-\frac{c_2}{2}\dot{\pi}^2-c_3\dot{\pi}^2\ddot{\pi}+\frac{3}{2}c_4\dot{\pi}^3\left(3H^2\dot{\pi}+2\dot{H}\dot{\pi}+8H\ddot{\pi}\right),\\
\label{eq:KG}
\beta\rho_m&=&-c_2\left(3H\dot{\pi}+\ddot{\pi}\right)+3c_3\dot{\pi}\left(3H^2\dot{\pi}+\dot{H}\dot{\pi}
+2H\ddot{\pi}\right)\nonumber\\
&& -18c_4H\dot{\pi}^2\left(3H^2\dot{\pi}+2\dot{H}\dot{\pi}+3H\ddot{\pi}\right).
\end{eqnarray}
The effective equation of state $w_{eff}$, equation of state $w_{\pi}$ and Hubble parameter for galileon model have the form
\begin{eqnarray}
\label{eq:rhopi}
w_{eff}&=&-\frac{2}{3}\frac{\dot{H}}{H^2}-1,\\
w_{\pi}&=&\frac{p_{\pi}}{\rho_{\pi}},\\
\label{eq:Galileon}
H^2&=& \frac{8 \pi G}{3} \left\lbrack \rho_{0m}(1+z)^3+\rho_{\pi}\right\rbrack,
\end{eqnarray}
where,
\begin{eqnarray*}
p_{\pi}&=&\frac{c_2}{2}\dot{\pi}^2+c_3\dot{\pi}^2\ddot{\pi}-\frac{3}{2}c_4\dot{\pi}^3\left(3H^2\dot{\pi}+2\dot{H}\dot{\pi}+8H\ddot{\pi}\right),\\
\rho_{\pi}&=&\frac{c_2}{2}\dot{\pi}^2-3c_3 H\dot{\pi}^3+\frac{45}{2}c_4 H^2\dot{\pi}^4.\\
\end{eqnarray*}
\begin{figure*}
\centering
\begin{center}
$\begin{array}{c@{\hspace{0.4in}}c}
\multicolumn{1}{l}{\mbox{}} &
        \multicolumn{1}{l}{\mbox{}} \\ [0.0cm]
\epsfxsize=2.5in
\epsffile{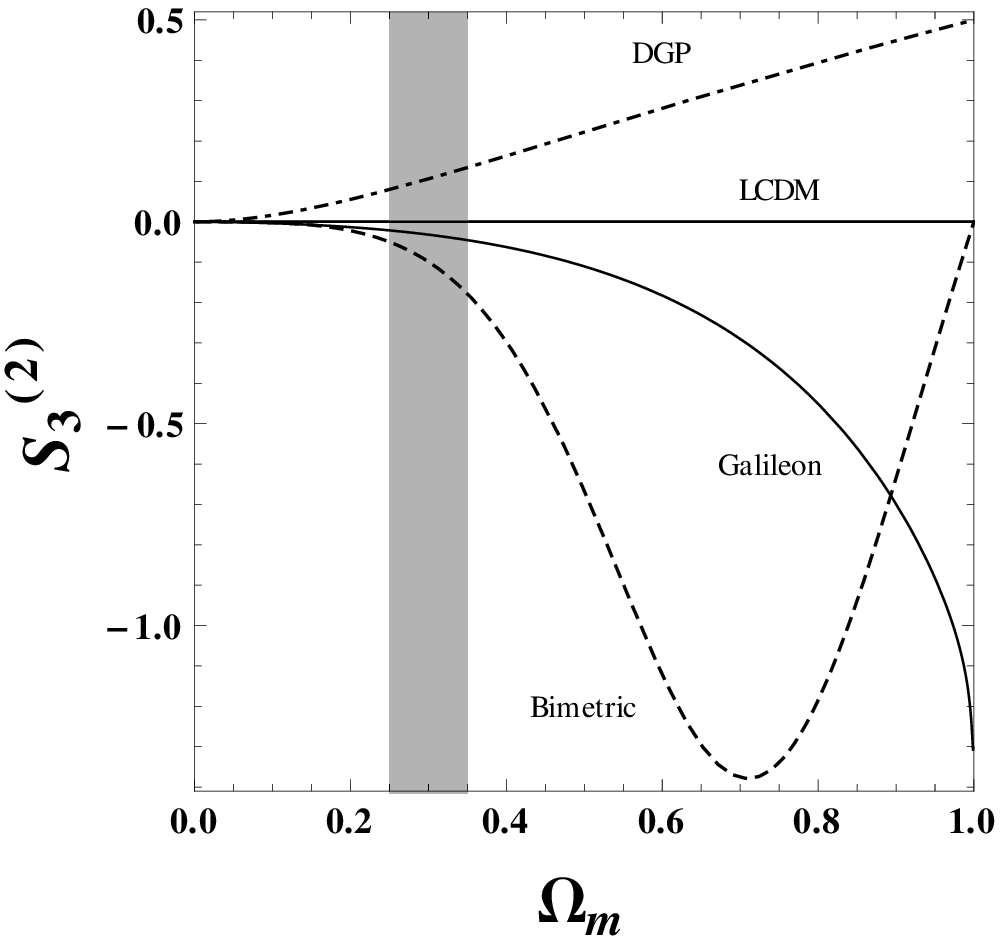} &
        \epsfxsize=2.44in
        \epsffile{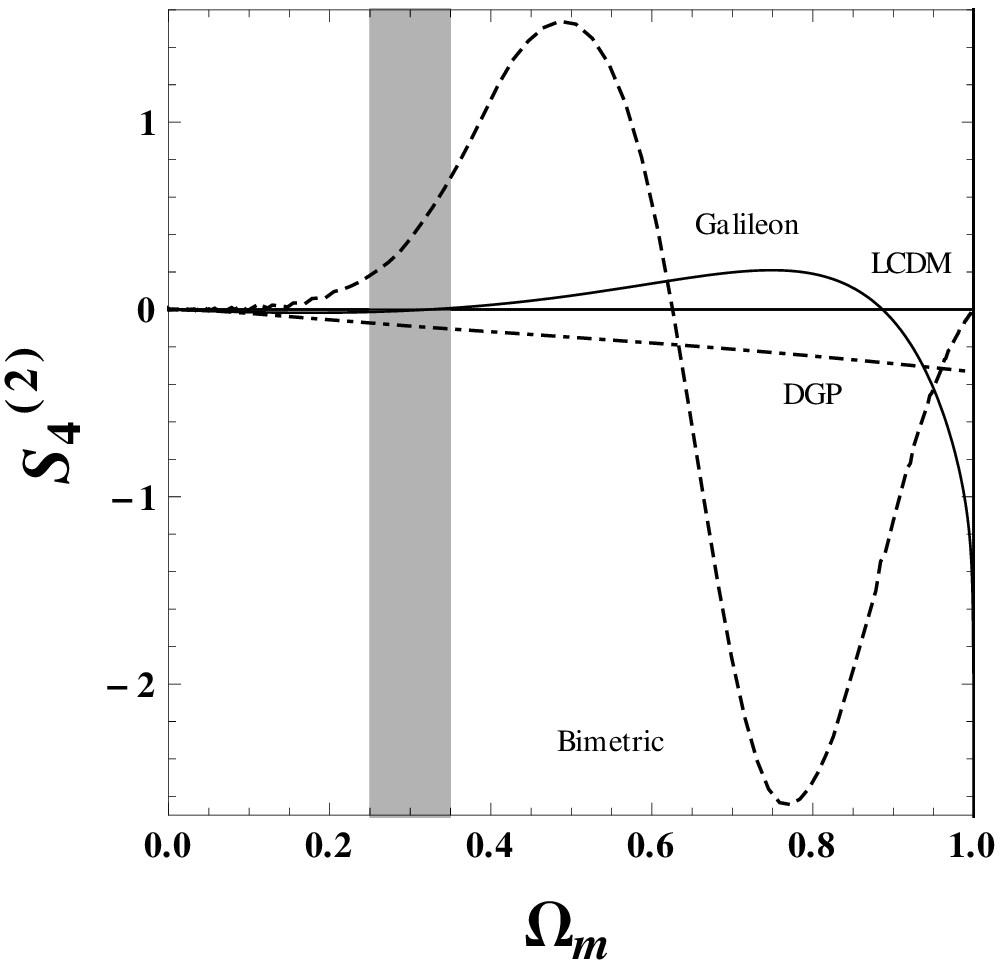} \\ [0.20cm]
\mbox{\bf (a)} & \mbox{\bf (b)}
\end{array}$
\end{center}
\caption{ \small The panels (a), (b) show the
Statefinder $S_3^{(2)}$, $S_4^{(2)}$ plotted against $\om$.
The dark energy models are: bimetric (dashed line), galileon
(solid line), DGP (dotdashed line). The horizontal black line shows LCDM. The vertical band centered at $\omm = 0.3$ roughly corresponds to the present epoch.}
\label{s32s42}
\end{figure*}
\begin{figure}
   \begin{center}
     \includegraphics[scale=0.65]{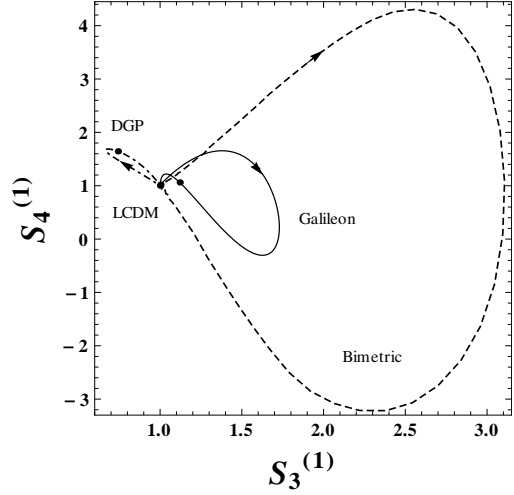}
 \end{center}
 \caption{ \small The Statefinders $S_4^{(1)}$ and $S_3^{(1)}$ are shown for different dark energy models; the arrows and dots show time evolution and present epoch respectively.
LCDM corresponds to fixed point (1,1). We assumed, $\omm = 0.3$ (present epoch).}
\label{s41s31}
\end{figure}
\begin{figure}
\begin{center}
\begin{tabular}{c c}
{\includegraphics[width=2.6in,height=2in,angle=0]{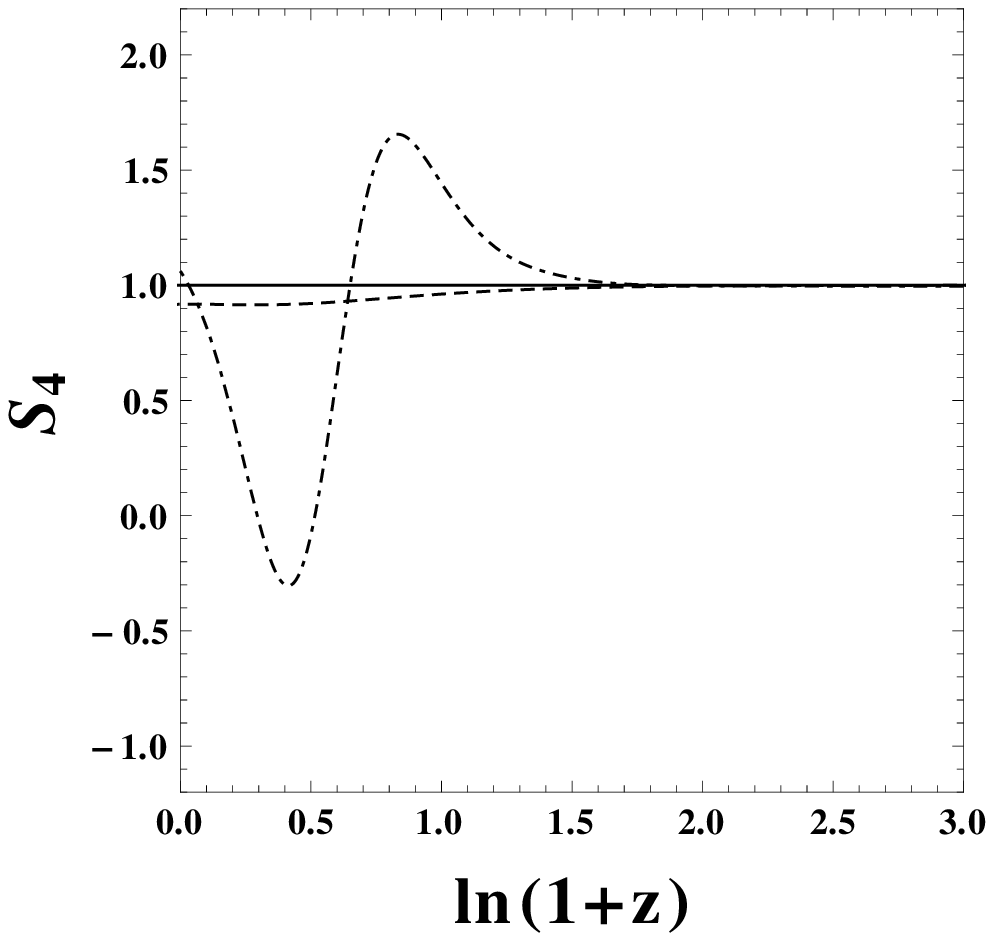}}&
{\includegraphics[width=2.6in,height=2in,angle=0]{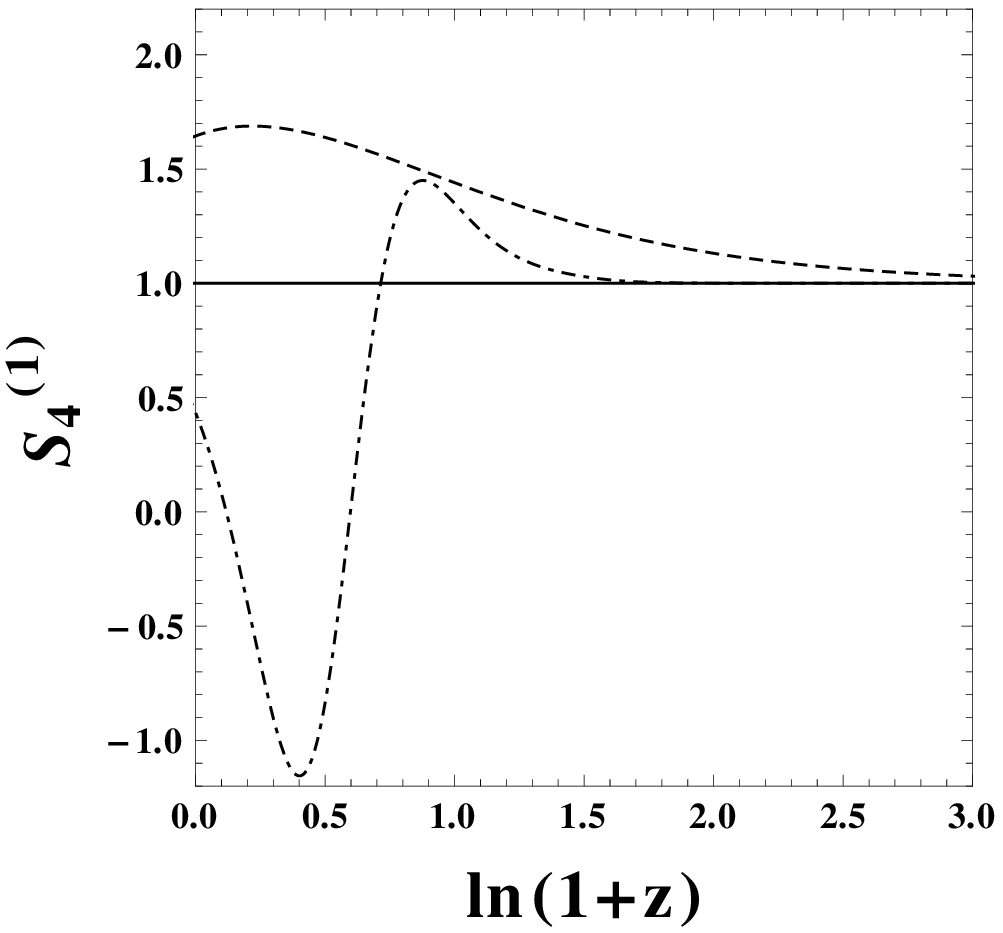}}
\\
{\includegraphics[width=2.6in,height=2in,angle=0]{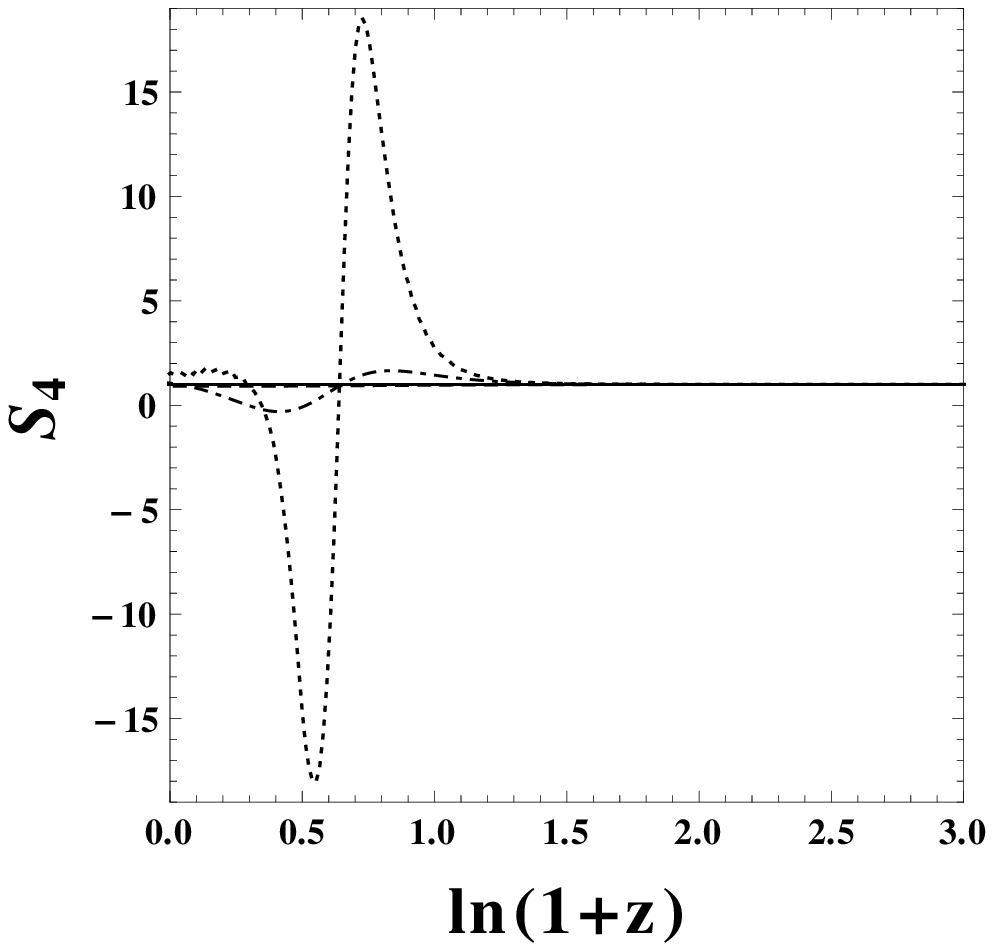}}&
{\includegraphics[width=2.6in,height=2in,angle=0]{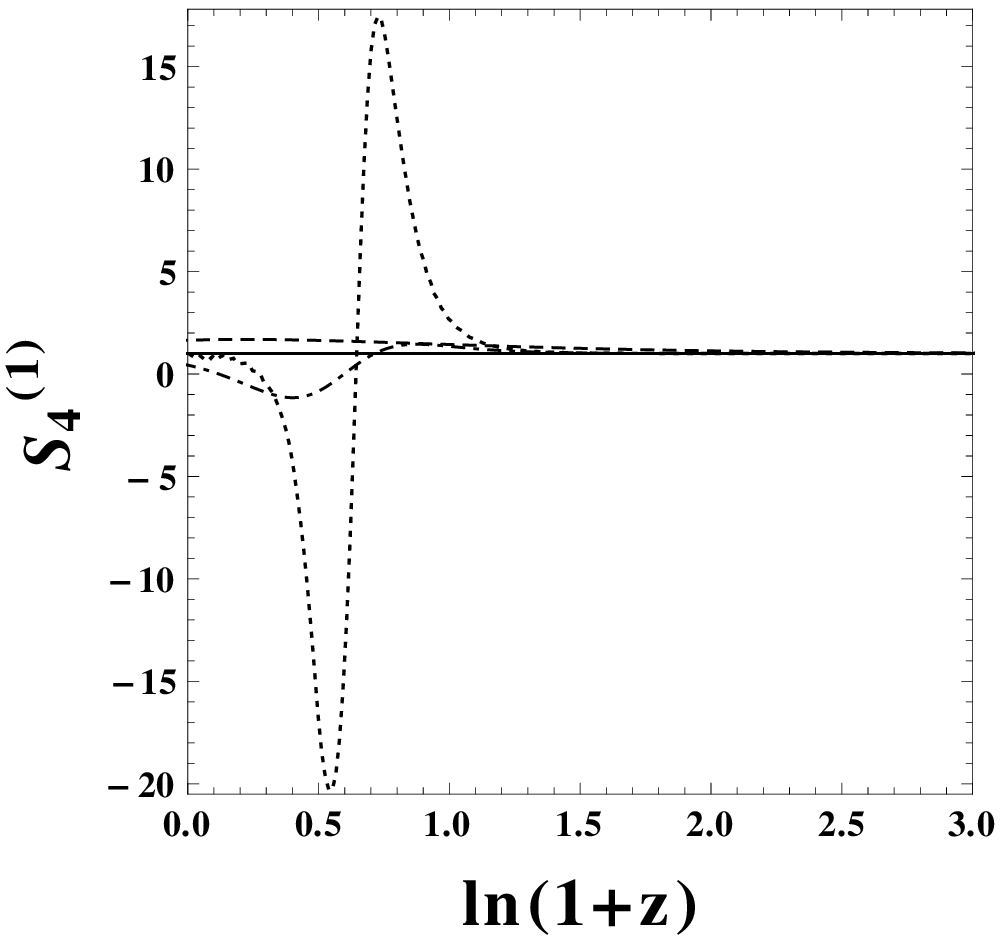}}
\end{tabular}
\caption{ This figure shows the evolution of  $ S_4$ and $S_4^{(1)}$ versus redshift z. For $\Lambda$CDM, $S_4^{(1)} := S_4 = 1$, and for other dark energy models the two definitions  $ S_4$ and $S_4^{(1)}$ have different values. Top panels show $\Lambda$CDM (solid line), galileon (dotdashed line) and DGP  models (dashed line) while bottom panels show the above models as well as the bimetric model (dotted line). From the top panel, one notes that $S_4^{(1)}$ provides a somewhat better diagnostic than $ S_4$.}
\label{s4z41znew}
\end{center}
\end{figure}
\begin{figure}
\begin{center}
\begin{tabular}{c c}
{\includegraphics[width=2.6in,height=2in,angle=0]{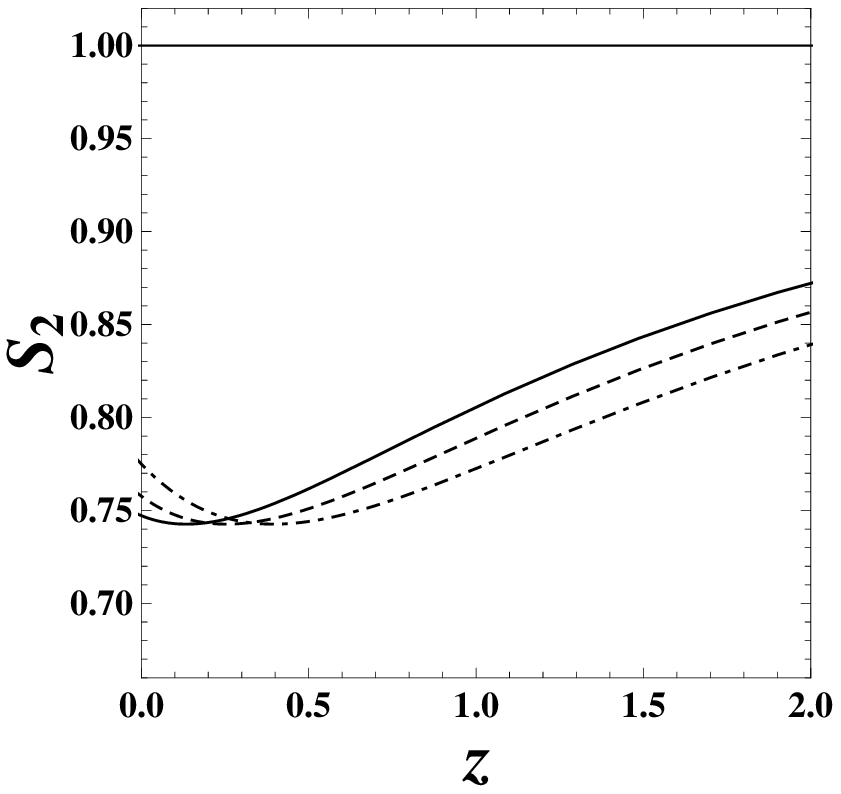}}&
{\includegraphics[width=2.6in,height=2in,angle=0]{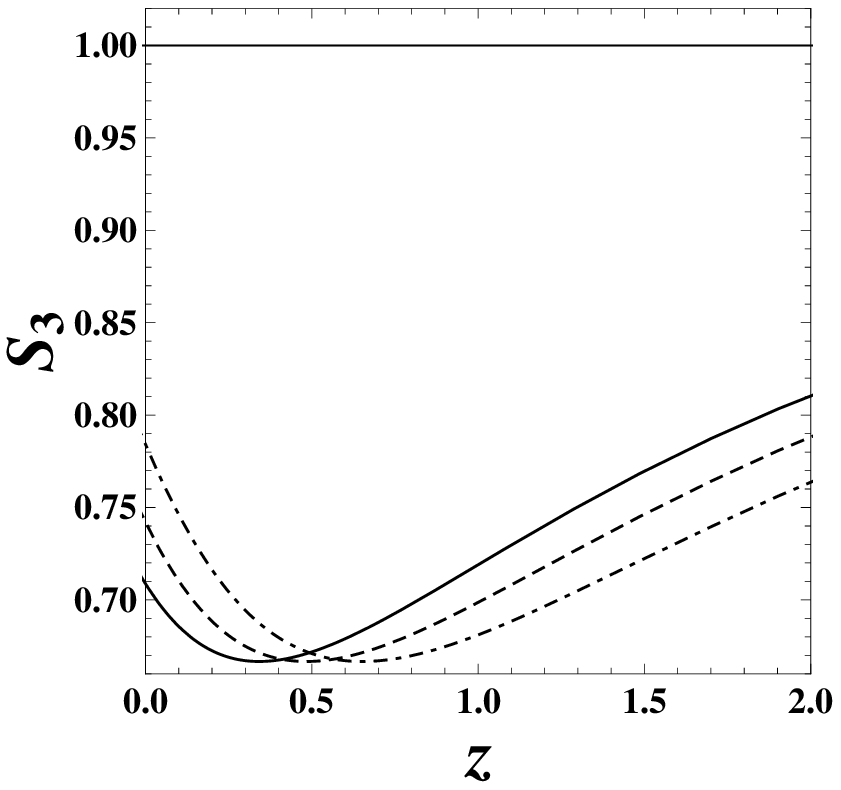}}
\\
{\includegraphics[width=2.6in,height=2in,angle=0]{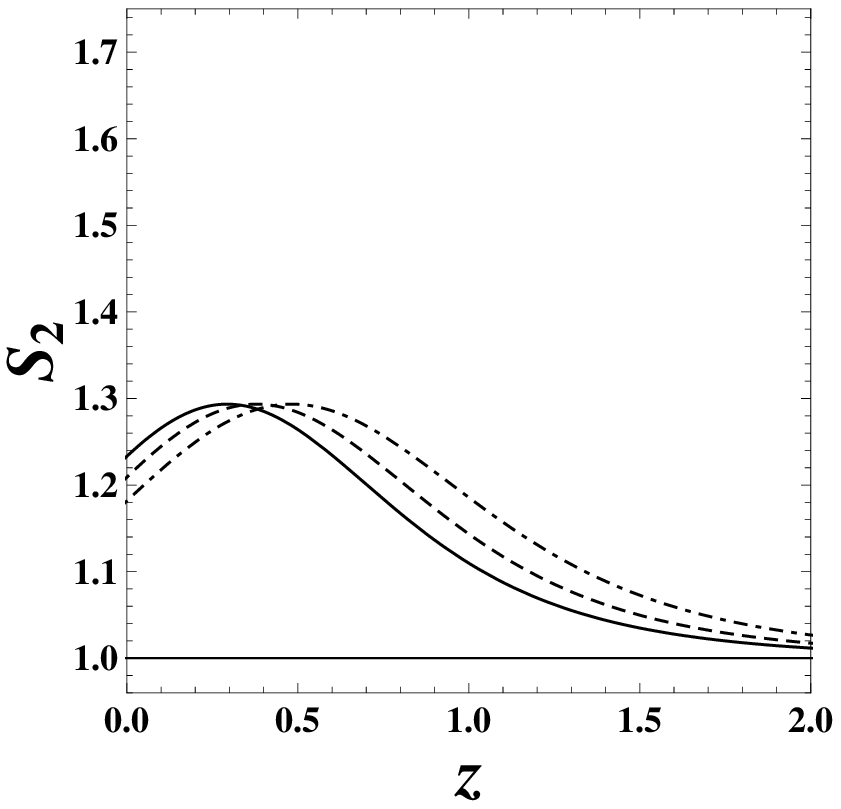}}&
{\includegraphics[width=2.6in,height=2in,angle=0]{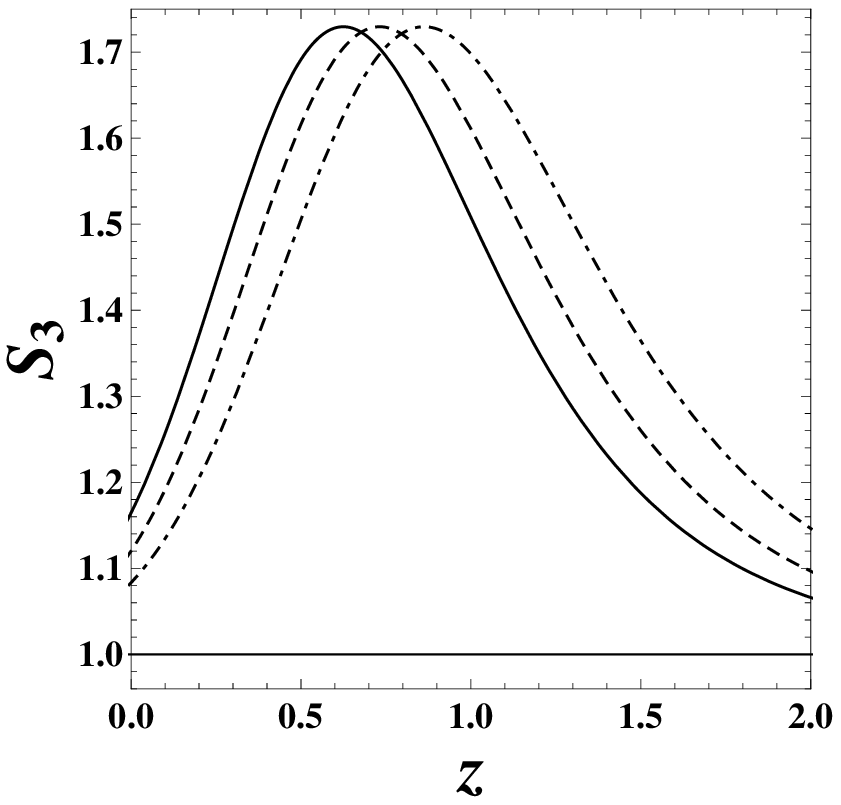}}
\\
{\includegraphics[width=2.6in,height=2in,angle=0]{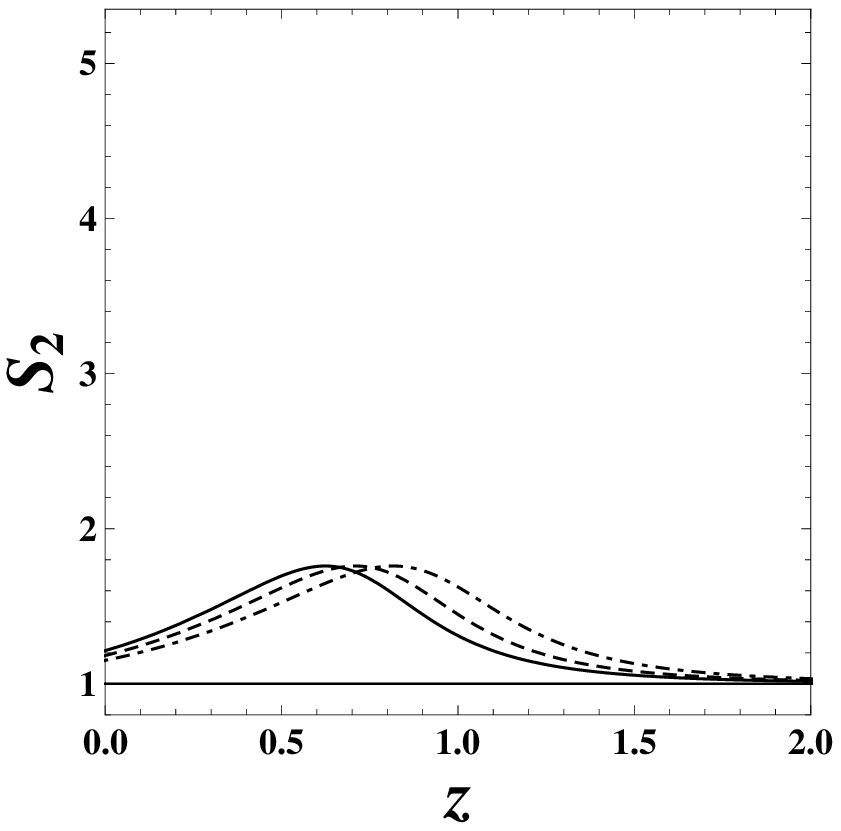}}&
{\includegraphics[width=2.6in,height=2in,angle=0]{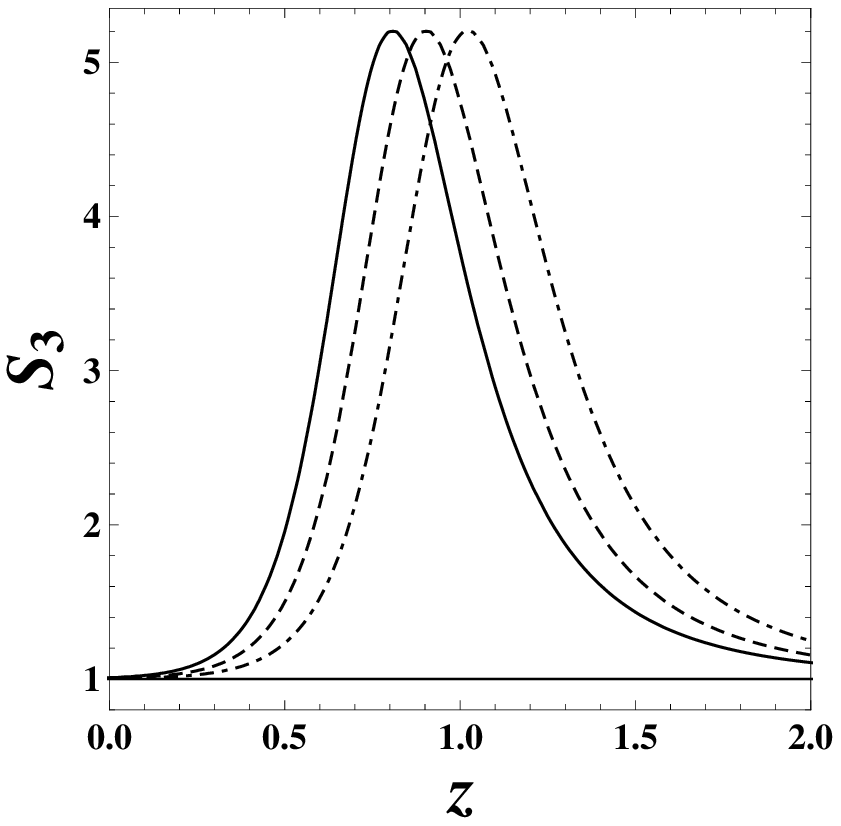}}
\end{tabular}
\caption{ Top, middle and bottom panels are for DGP, galileon and bimetric models respectively. In all panels, $\omm$=0.25 (dotdashed line), $\omm$=0.3 (dashed line), $\omm$=0.35 (solid line) and $\lam$CDM (horizontal solid line). The effect of $\omm$ on $S_3$ is more severe than $S_2$ as  $S_3$ has cube power of $H$ in the denominator.}
\label{s2zs3z}
\end{center}
\end{figure}
\item {\bf The DGP model} \citep{dgp}:
\begin{equation}
\label{eq:DGP}
\frac{H(z)}{H_0} = \left[ \left(\frac{1-\omm}{2}\right)+\sqrt{\omm (1+z)^3+
\left(\frac{1-\omm}{2}\right)^2} \right]\,\,
\end{equation}
\end{itemize}
In order to apply the statefinder analysis to the models under
consideration, we notice that the deceleration parameter q,
statefinder pair $\statei$ and snap $\alpha_4$ can be easily
expressed in terms of Hubble parameter $H(z)$ and its derivatives
 as follows:
\begin{eqnarray}
\label{eq:snap}
q(z)&=&(1+z) \frac{H'}{H} - 1 ,\nonumber \\
r(z)&=&1-2(1+z) \frac{H^{\prime}}{H}+(1+z)^2 \left(\frac{H^{\prime\prime}}{H}+\frac{H'^2}{H^2}\right),\nonumber \\
s(z)&=&\frac{r(z)-1}{3 ( q(z)-1/2)},\nonumber\\
\alpha_4(z)&=&1-3(1+z) \frac{H'}{H}+(1+z)^2\left(\frac{H''}{H}+3\frac{H'^2}{H^2}\right)\nonumber \\
&& -(1+z)^3 \left(\frac{H'''}{H} +4 H'\frac{H''}{H^2}+\frac{H'^3}{H^3}\right),
\end{eqnarray}
where, $H$ is given by (\ref{eq:bm}), (\ref{eq:Galileon}) and (\ref{eq:DGP}) for different dark energy models discussed in the literature.

We are now in a position to present the numerical results for
statefinder parameters for each model and strike a comparision
between them. In figure \ref{rs-rq}, we show the time evolution of the
statefinder pairs $\statei$ and $\statej$ for bimetric model of
massive gravity and DGP model. One can see that both the models are
differentiated by statefinder pairs. The statefinder hierarchy is
used for higher derivatives of scale factor; which can easily
distinguish
different dark energy models. In figure \ref{s2s31}, the behavior of  bimetric, galileon and DGP models show nondegeneracy around the
present value of $\om$  in $S_2-\om$ and $S_3-\om$ plane.

In figure \ref{s41}, galileon model is highly degenerate, DGP model is nearly  degenerate whereas bimetric model of massive gravity is showing
nondegeneracy around the present value of $\om$. In $S_3^{(2)}-\om$ plane, galileon and bimetric models are nearly degenerate whereas DGP model is  nondegenerate, around the present value of $\om$, as shown in figure \ref{s32s42} (a). In $S_4^{(2)}-\om$ plane, galileon model is highly degenerate, DGP model is nearly  degenerate whereas bimetric model of massive gravity is nondegenerate, around the present value of $\om$, as shown in figure \ref{s32s42} (b). In figure \ref{s41s31}, we exhibit the phase portrait in $S_4^{(1)}-S_3^{(1)}$ plane, where degeneracies among different dark energy
models are broken.

Particularly, we should comment on  $S_2-\om$, $S_3-\om$ and $S_4^{(1)}-S_3^{(1)}$ planes. From figures \ref{s2s31} and \ref{s41s31}, one can clearly see that both the figures do not have degeneracies among $\Lambda$CDM, bimetric, galileon and DGP models around the present value of $\om$.
At higher redshift (corresponding to $ \om \simeq 0.7$), $S_3^{(2)}$
does an excellent job in distinguishing between the different DE
models, as shown in figure \ref{s32s42} (a). Other statefinder
hierarchy does not seem to perform well in distinguishing the
different dark energy models considered in this paper. While
carrying out comparison  between $S_2$ and $S_3$, we note that
Statefinder $S_2$ is easier to determine as it requires only two
derivatives of the expansion factor whereas $S_3$ requires three.
Since $S_2$, explicitly  depends upon $\Omega_m$, it might look at
the onset that it would exhibit more sensitivity to $\omm$ than
$S_3$. However, we should keep in mind that  both $S_2$ and $S_3$
also  implicit dependence on $\omm$ through Hubble parameter; they
contain $H^2$ and $H^3$ terms in the denominator respectively.
 Indeed, our numerical simulation shows that uncertainties in
 the value of  $\omm$  affect $S_3$ more severely
 than $S_2$, as demonstrated in figure \ref{s2zs3z},
 where $S_2$ and $S_3$ are plotted against redshift z.
  But when  $S_2$ and $S_3$ are plotted against $\om$,  there are no
  effects on $S_2$ and $S_3$ for different values
  of $\omm$, as shown in figure \ref{s2s31}
  (though this figure is plotted for $\omm = 0.3$,
  it is checked that the results are same for other
  values). An important comment on the relative matter density dependence of $S_2$ and $S_3$ is in order. We know that $S_3$ depends solely on the third derivative of the scale factor $a(t)$ and its value can be determined from observations of the luminosity distance $D_L$ or the Hubble parameter, after differentiation. It is important to note that  the value of the matter density does not enter into the expression of $S_3$. Therefore, if the observed value of $S_3$ departs from unity,  it will give an important information about the nature of dark energy, namely, that dark energy is something beyond cosmological constant. On the other hand, a similar argument does not hold for $S_2$ since its value explicitly depends upon $\om$ and so the latter needs to be known (from observation) even after the differentiation of $D_L$ of $H(z)$.
  
It is interesting to note that Planck results favor larger values of $ \omm $ than SN$1$a and this is well reflected in the vertical band around $\omm \simeq 0.3$ in most of the figures. The fact that an incorrect value of matter density could significantly bias the reconstructed value of $\omega(z)$ was discussed in \cite{alamwz}.
\section{Conclusion}
\vspace{3mm}

In this paper we have shown that the bimetric model of massive
gravity and the DGP model can be distinguished  by using the
statefinder pairs $\statei$ and $\statej$. We also carried out
comparison between bimetric theory of massive gravity, galileon
modified gravity and other popular dark energy models using the
statefinder hierarchy in concordance cosmology.  Our investigation
in $S_2-\om$ and $S_3-\om$ planes show nondegeneracy among $\Lambda$CDM,
bimetric, galileon and DGP models around the present value of $\om$,
and all models considered in this paper are successfully
differentiated by statefinder hierarchy on these planes. We have also
noticed that  our analysis presents a good comparison among the
different dark energy models considered in this paper. Figures
\ref{s2s31} and \ref{s41s31} show nondegeneracy among
popular dark energy models around the present value of $\om$. We
find that $S_4^{(1)}$, $S_3^{(2)}$ and $S_4^{(2)}$ do not perform
well in distinguishing the different dark energy models considered
in this paper, as shown in figures \ref{s41} and \ref{s32s42} . By looking at the success of the statefinder hierarchy diagnostic, we are encouraged
to apply the analysis to models of extended massive gravity
discussed in the literature.

We have demonstrated that $S_2$ and $S_3$ perform better than
the higher order statefinders in discriminating between $\Lambda$CDM
and modified gravity models of dark energy. While comparing  between
$S_2$ and $S_3$, we find that the statefinder $S_2$ is better
discriminant than $S_3$, as demonstrated in figure \ref{s2s31}. A
comment about the possibility of observational constraints on the
statefinder parameters is in order. Since these parameters include
higher time derivatives, it is really challenging to measure them
with good accuracy. One must wait until Euclid for a precise
determination of this diagnostic, which is a very exciting
possibility, since Euclid is most likely to take off and the data
should be available within about a decade. Meanwhile as demonstrated
in Ref. \cite{statefinder}, one can use mean statefinder statistics
using SNAP type experiment to distinguish various models. Other mean
diagnostics are discussed in Refs. \cite{Om, wbar}
\section*{Acknowledgment}
The authors are grateful to S. Deser for his useful comments and for bringing Refs. \cite{sdg, sdmg} to our attention.
We thank M. Sami, V. Sahni, S. Jhingan and Gaveshna Gupta for useful
discussions, and also thanks to Sarita Rani for helping us in improving this manuscript. MS acknowledges the financial support provided by the
DST, Govt. of India, through the research project No.
SR/S2/HEP-002/2008.

\end{document}